\documentclass[prl,aps,showpacs,twocolumn]{revtex4}
\usepackage{graphicx}

\begin{document}

\title{Evidence for resonant scattering of electrons by spin fluctuations in  $LaNiO_3/LaAlO_3$ heterostructures grown by pulsed laser deposition}

\author{S. Sergeenkov$^{1}$,  L. Cichetto, Jr.$^{2,3,4}$, E. Longo$^{3,4}$ and  F.M. Araujo-Moreira$^{2}$}

\affiliation{$^{1}$ Departamento de F\'{i}sica, CCEN, Universidade Federal da Para\'{i}ba,  58051-970 Jo$\tilde{a}$o Pessoa, PB, Brazil\\
$^{2}$ Departamento de F\'{i}sica, Universidade Federal
de S$\tilde{a}$o Carlos, 13565-905 S$\tilde{a}$o Carlos, SP,  Brazil\\
$^{3}$ LIEC - Department of Chemistry, Universidade Federal de S$\tilde{a}$o Carlos,  13565-905  S$\tilde{a}$o Carlos, SP, Brazil\\
$^{4}$ Institute of Chemistry, Universidade Estadual Paulista - Unesp, 14801-907 Araraquara, SP, Brazil}

\date{ \today}

\begin{abstract}
We present measurements of resistivity $\rho$ in highly oriented $LaNiO_3$ films grown on $LaAlO_3$ substrates by using a pulsed laser deposition technique. The experimental data are found to follow a universal $\rho (T) \propto T^{3/2}$ dependence for the entire temperature interval ($20K<T<300K$). The observed behavior has been attributed to a resonant  scattering of electrons on antiferromagnetic fluctuations (with a characteristic energy $\hbar \omega _{sf}\simeq 2.1meV$) triggered by spin-density wave propagating through the interface boundary of $LaNiO_3/LaAlO_3$ sandwich. 
\end{abstract}


\maketitle

{\bf 1. Introduction.} 
Even though $LaNiO_3$ (LNO) belongs to the nickelates family $RNiO_3$ (with $R$ being a rare-earth element), it possesses unique physical properties because it does not undergo a metal-insulator transition (MIT) from paramagnetic (PM) metal to antiferromagnetic (AFM) insulator like all the other members of this family. It stays PM metal for all temperatures. However, recent investigations (see, e.g., [1-5] and further references therein) suggest that one can significantly modify LNO properties by depositing it on different substrates. This can be achieved by thickness controlled partial suppression of the charge ordering (which is believed to be responsible for manifestation of MIT in nickelates). The most interesting result of such a heterostructure engineering is probably the one that leads to appearance of a new magnetic structure, the so-called pure spin-density wave (SDW) material exhibiting properties of an AFM metal [6-11]. Such heterostructures are found to manifest very unusual properties (both magnetic and transport related). In particular, it has been successfully proven experimentally [10,11] that depositing LNO thin film on $LaAlO_3$ (LAO) substrate results in appearance of a rather robust AFM order inside LNO/LAO superstructure (with Neel temperature $T_N\simeq 100K$) due to formation of SDW and concomitant charge redistribution in LNO films [7]. 
In this Letter we present our latest measurements of resistivity in $(l00)$-oriented LNO thin films deposited on $(l00)$-oriented LAO substrate by using the pulsed laser deposition (PLD) technique. We demonstrate that our data can be very well fitted by a rather simple law $\rho (T)=A+BT^{3/2}$ for the entire temperature interval  ($20K<T<300K$). We argue that such a temperature dependence is a result of strong resonant scattering of conducting electrons on thermally excited AFM spin fluctuations (taking place inside LNO/LAO hybrid structure) which completely suppresses all the other scattering mechanisms (such as electron-phonon and electron-electron interactions).

{\bf 2. Experimental.} Recall that PLD technique is especially suitable for providing high quality samples with atomically smooth surfaces whose composition agrees well with that of the target and the films can be deposited at a wide range of oxygen pressure. Therefore, PLD technique was employed to deposit thin films of LNO on $(100)$ oriented LAO substrate with typical dimensions of $5\times 5 \times 0.5 mm^{3}$. Laser wavelength and repetition rate were $\lambda = 248 nm$ ($KrF$ laser with $25 ns$ pulse duration) and $f = 2 Hz$, respectively.  

Microstructure and crystallographic orientation of the films were characterized by X-ray diffraction (XRD) scans. The surface morphology was confirmed by atomic force microscopy and scanning electron microscopy (these results and further details will be presented elsewhere). XRD pattern of the discussed here LNO/LAO hybrid structure (with LNO thickness of $t=60nm$) is depicted in Fig.1. 

The electrical resistivity $\rho (T)$ was measured using the conventional four-probe method. To avoid Joule and Peltier effects, a dc current $I=100\mu A$ was injected
(as a one second pulse) successively on both sides of the sample. The voltage drop $V$ across the sample was measured with high accuracy by a $KT256$ nanovoltmeter.
\begin{figure}
\centerline{\includegraphics[width=7.50cm, angle=270]{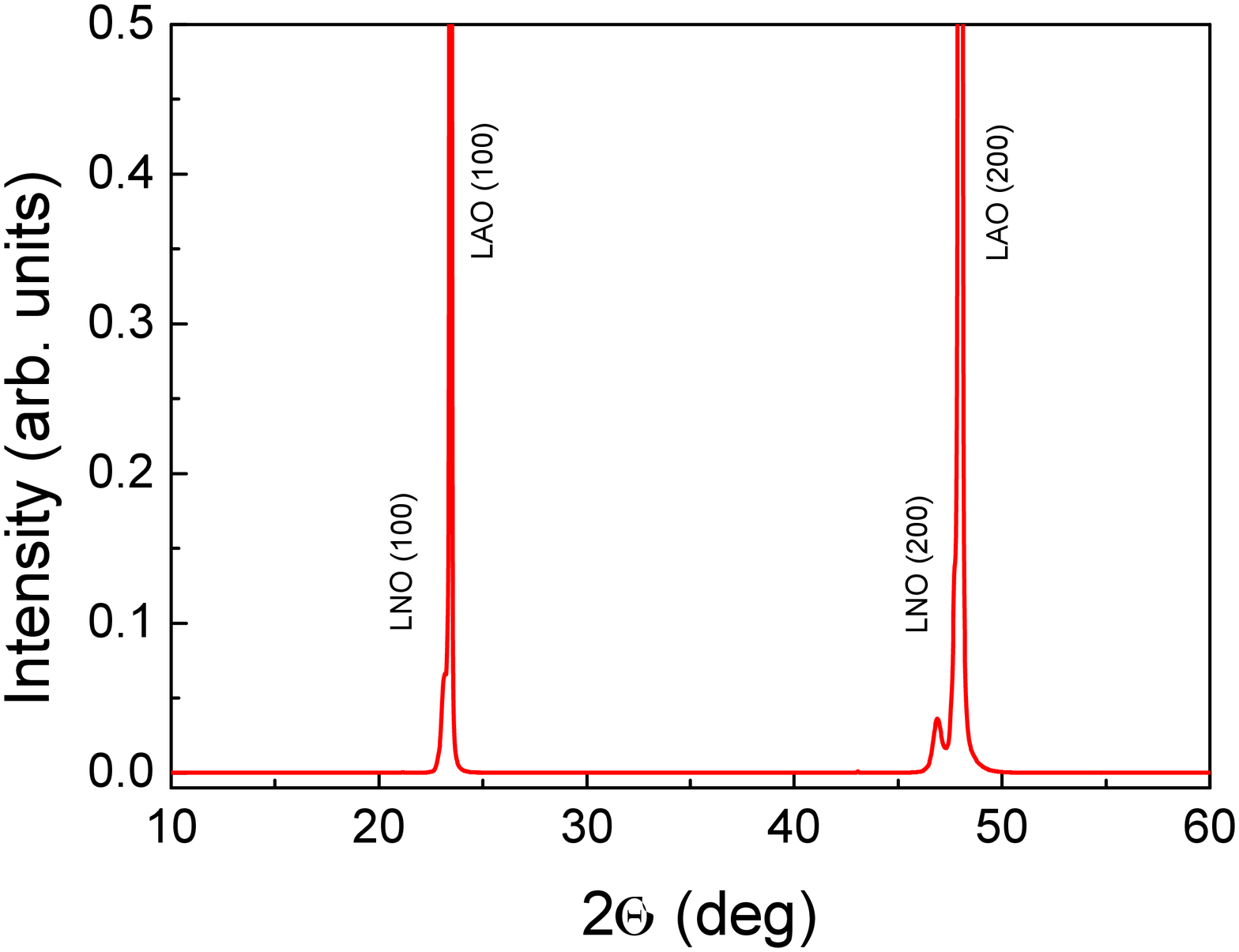}}
\vspace{0.5cm} \caption{X-ray diffraction
spectrum of LNO films deposited on oriented LAO substrate. } 
\end{figure}
\begin{figure}
\centerline{\includegraphics[width=7.50cm]{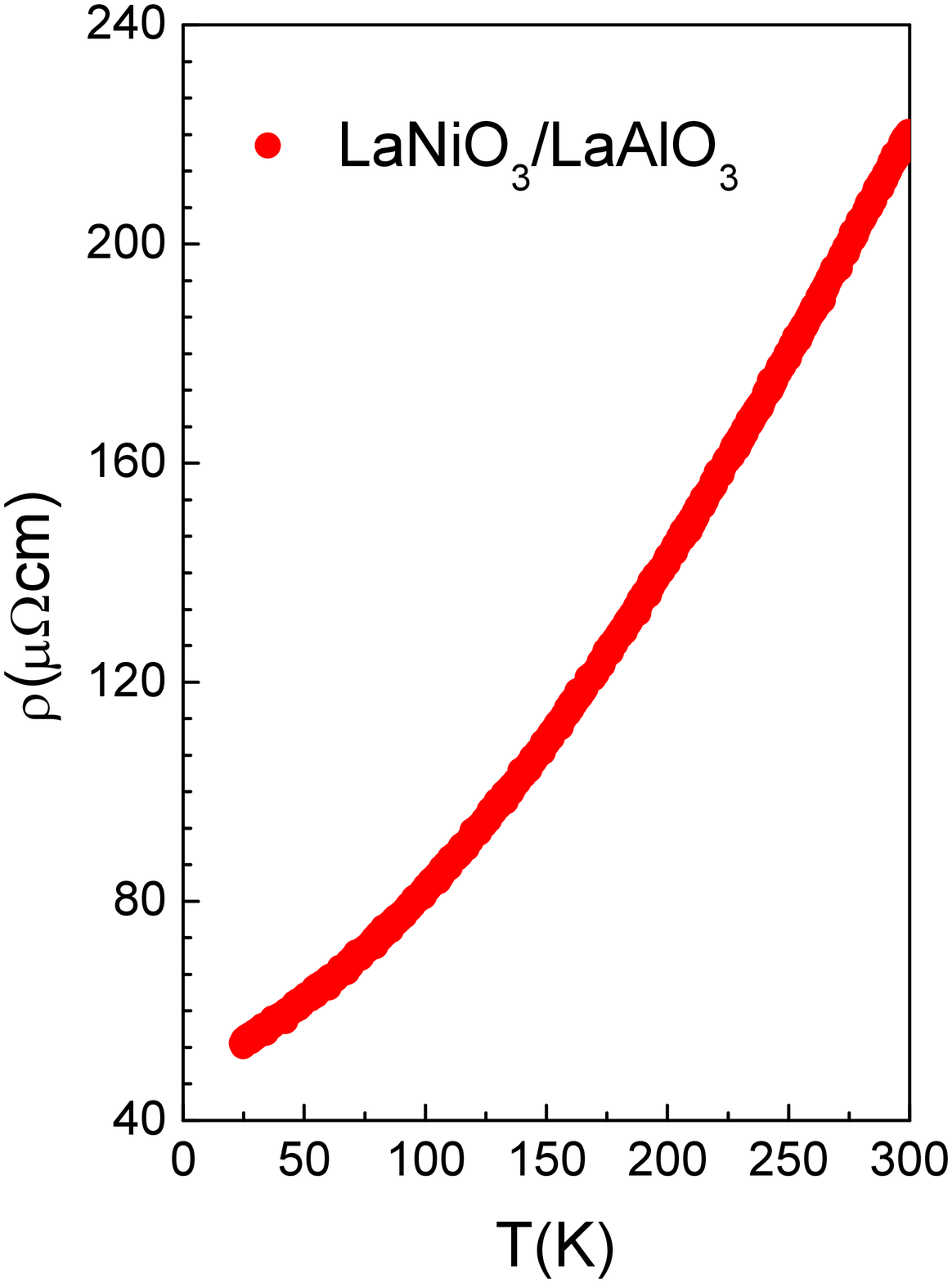}}\vspace{0.5cm}
\caption{ Temperature dependence of the
resistivity $\rho (T)$ measured for a typical $LaNiO_3$ thin film deposited on oriented $LaAlO_3$.}
\end{figure}
\begin{figure}
\centerline{\includegraphics[width=7.50cm]{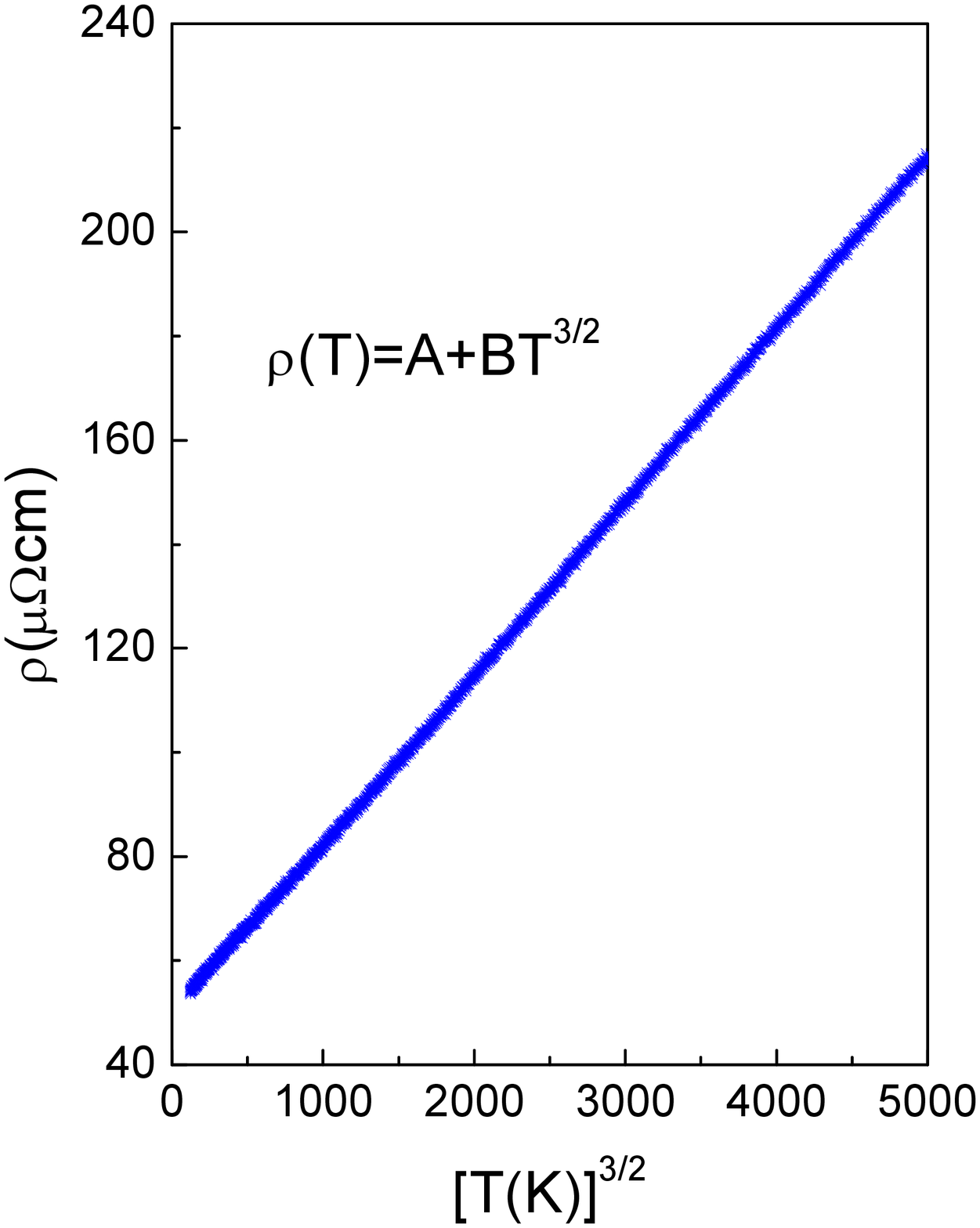}}\vspace{0.5cm}
\caption{The best fit of the experimental data shown in Fig.2 according to Eq.(4).}
\end{figure}

{\bf 3. Results and Discussion.} 
Fig.2 shows the typical results for the temperature
dependence of the resistivity $\rho (T)$ in our $LaNiO_3/LaAlO_3$  thin films  heterostructure. Fig.3 presents the best fit of the experimental data for \textit{all} temperatures according to the following fitting expression: $\rho (T)=A+BT^{3/2}$ with $A=50\mu \Omega cm$ and $B=0.0033\mu \Omega cm/K^{3/2}$. 
Given the above discussion on appearance of AFM order in LNO/LAO hybrid structure, it is quite reasonable to assume that the observed temperature behavior can be attributed to the manifestation of strong long-range AFM spin fluctuations (with a characteristic energy $\hbar \omega _{sf}\simeq k_BT_{sf}$ corresponding to low-energy spin dynamics spectrum measured by inelastic neutron scattering experiments). 

It should be mentioned that the origin of the overwhelming $T^{3/2}$ dependence due to spin fluctuations in AFM metals with nontrivial topology of the Fermi surface has been discussed before within a detailed microscopic picture [12,13]. In particular, it has been demonstrated (within a rigorous Boltzmann approach) that [12] in weakly disordered metals close to AFM quantum-critical point ($p_c$) the anisotropic scattering from critical spin fluctuations is strongly influenced by weak (but isotropic) scattering from small amounts of disorder, resulting in a scaling like dependence of magnetoresistivity valid for all temperatures:

\begin{equation}
\rho (T) \propto T^{3/2}F\left[T/\rho_0, (p-p_c)/\rho_0, H/ \rho_0^{3/2}\right]
\end{equation}
where $F$ is a scaling function with $\rho_0$ the residual contribution (due to impurity scattering), $H$ the magnetic field, and $p-p_c>0$ measuring the distance from the quantum-critical point (QCP) on the paramagnetic side of the phase diagram. 

A more detailed analysis revealed [12] that at zero magnetic field a 3D system (single crystal) with a small amount of disorder exhibits a resistivity crossover (with increasing  the temperature) from $T^{3/2}$ (at low temperatures) to $T$ behavior (at higher  temperatures). However, if the system is tuned to the QCP in a finite field, $\rho(T) \propto T^{3/2}$ dependence is expected to be observed at higher temperatures as well. More precisely, it was found [12] that at high enough magnetic fields, the temperature dependence of $\rho (T)$ actually saturates at $T^{3/2}$ at the highest temperatures. 

At the same time, a similar $T^{3/2}$ dependence of resistivity has been observed in  the high-pressure paramagnetic state of 3D metals and has been attributed to manifestation of small-angle \textit{interband} inelastic scattering by AFM spin fluctuations (Cf. Eq.(9) from Ref.[13]). Moreover, to explain the universality of the observed $T^{3/2}$ behavior over the entire measured temperatures, it was suggested [13] that the scattering by \textit{intraband} fluctuations replenishes the electron distribution (depleted by \textit{interband} inelastic scattering) while having a little effect on the current, thus making \textit{intraband} scattering mechanism responsible for the robustness of the $T^{3/2}$ behavior in deformed 3D metal.  

Based on the above observations and given the well-known fact that thin enough 2D films can be effectively treated as 3D systems under high enough pressure (or equivalent magnetic field), it is quite reasonable to assume that the experimentally observed behavior of resistivity in our "dirty" films can indeed be dominated by AFM spin fluctuations for the entire temperature interval. 
For simplicity, in this paper we adopt a phenomenological approach based on our previous experience in superconductivity [14]. More specifically, to describe fluctuations induced thermal broadening effects, we suggest the following scenario. Namely, we introduce the temperature dependence via the cutoff frequency $\Omega (T)=U(T)/\hbar$ which accounts for AFM fluctuations with an average thermal energy $U(T)=\frac{1}{2}C<u^2> \simeq k_BT$ where $C$ is the force constant of a two-dimensional harmonic oscillator, and $<u^2>$ is the mean square displacement of the magnetic atoms from their equilibrium positions.
As a result, we arrive at the following simple expression for the temperature dependence of resistivity in our hybrid film structure [14]
\begin{equation}
\rho (T)=\rho_0+\rho_{sf}\left[1+\int_{\omega _{sf}}^{\omega _{sf} +\Omega (T)
 } d\omega f(\omega)\right]
\end{equation}
where $\rho (0)=\rho_0+\rho_{sf}$ is the total residual contribution, and $f(\omega)$ is an appropriate spectral distribution function.

It should be pointed out that in order to correctly model SDW created fluctuations spectrum in our heterostructure, instead of the previously used Drude-Lorentz law (which is quite appropriate for treating spin fluctuations in superconducting films [14]), we make use of another spectral distribution (which is more suitable for treating spin fluctuations in SDW type materials) valid for $\omega _{sf}\le \omega \le \omega _{sf}+\Omega (T)$:
\begin{equation}
f(\omega)=\frac{2}{\pi \omega _{sf}^2}\sqrt{\omega ^2-\omega _{sf}^2}
\end{equation}
A careful analysis of Eqs.(2) and (3) reveals that the seeking $T^{3/2}$ law corresponds to a resonance like condition $\omega \simeq \omega _{sf}$ of the above distribution,  which becomes $f(\omega)\propto \sqrt{\omega -\omega _{sf}}$ under this condition. It can be directly verified now that the resonant like SDW governed spectrum indeed results in the observed dependence of the resistivity
\begin{equation}
\rho (T)=A+BT^{3/2}
\end{equation}
with 
\begin{equation}
B=\frac{4\sqrt{2}}{3\pi}\left(\frac{k_B}{\hbar \omega _{sf}}\right)^{3/2}\rho_{sf}
\end{equation}
and $A=\rho_0+\rho_{sf}$.

Notice that according to Eq.(5), the absolute value of the resonance frequency $\omega _{sf}$ in the spectrum of SDW driven AFM fluctuations (dominating the scattering mechanism) can be estimated if we know the value of the residual contribution $\rho_{sf}$ due to electron scattering by spin fluctuations, which is expected to be rather small in comparison with residual contribution $\rho_{0}$ due to electron scattering by impurities (very much like in superconducting films [14]). In turn, the latter contribution can be estimated as follows [14]: $\rho_0=1/\omega_p^{2}\epsilon_0 \tau_0$ where $\omega_p$ is the plasmon frequency, $1/\tau_0$ is the elastic scattering rate, and $\epsilon_0 = 8.85\times 10^{-12}F/m$ is the vacuum permittivity. Using [5] $\hbar\omega_p\simeq 1.1 meV$ and $1/\tau_0\simeq 1.2\times 10^{7}s^{-1}$ for our heterostructure, we obtain $\rho_0\simeq 40\mu \Omega cm$. Now, using the experimentally found value of $A=\rho_0+\rho_{sf}=50\mu \Omega cm$, we obtain  $\rho_{sf}=A-\rho_0\simeq 10\mu \Omega cm $ for the seeking estimate. With this value in mind and recalling that $B=0.0033\mu \Omega cm/K^{3/2}$, from Eq.(5) we can estimate now the value of the resonance frequency $\omega _{sf}$. The result is as follows, $\hbar \omega _{sf}\simeq 2.1meV$ which gives $T_{sf}\simeq 23K$ for the onset temperature where spin fluctuations begin to dominate the scattering process in our $LaNiO_3/LaAlO_3$  thin films  heterostructure, in a good agreement with the observations (Cf. Fig.2). However, in order to further check a plausibility of the presented here scenario and to corroborate the obtained here results, it is important to relate the resonant frequency $\omega _{sf}$ with the inelastic neutron scattering data on low-energy spin dynamics (for the energy spectrum ranging from $0.5meV$ to $5meV$) in such hybrid structures. It is interesting to notice that for our samples $\omega _{sf}\simeq 2\omega_p$.

And finally, it is worthwhile to mention that somewhat similar unusual magnetic properties have been observed in other topologically nontrivial systems, including quasi 1D conducting polymers with AFM ground state (where SDW like spin transport is mediated by topological solitons [15]) and highly ordered artificially prepared 2D arrays of Josephson junctions (exhibiting pronounced geometrical quantization effects [16]).

In summary, a universal $\rho (T) \propto T^{3/2}$ dependence of resistivity for the entire temperature interval was observed in $LaNiO_3$ thin films grown on oriented $LaAlO_3$ substrate (by using a pulsed laser deposition technique) and attributed to resonant scattering of conducting electrons on spin fluctuations resulting in thermally activated displacement of magnetic atoms. 

We are indebted to the Referee for very useful comments which helped us better understand the obtained results. This work was financially supported by Brazilian agencies FAPESQ (DCR-PB), FAPESP and CNPq. We are very thankful to CEPID CDMF 2013/07296-2 and FAPESP (process 2014/01371-5) for continuous support of our project on nickelates.

\end{document}